# Ferromagnetism and spin-dependent transport at a complex oxide interface

Yilikal Ayino[1], Peng Xu[2], Juan Tigre-Lazo[1], Jin Yue[2], Bharat Jalan[2], and Vlad S. Pribiag[1*]

1. School of Physics and Astronomy, University of Minnesota, Minneapolis, MN 55455, USA

2. Department of Chemical Engineering and Materials Science, University of Minnesota, Minneapolis, MN 55455, USA

*e-mail: vpribiag@umn.edu

**Abstract:**

Complex oxide interfaces are a promising platform for studying a wide array of correlated electron phenomena in low-dimensions, including magnetism and superconductivity. The microscopic origin of these phenomena in complex oxide interfaces remains an open question. Here we investigate for the first time the magnetic properties of semi-insulating $NdTiO_3$/$SrTiO_3$ (NTO/STO) interfaces and present the first milli-Kelvin study of NTO/STO. The magnetoresistance (*MR*) reveals signatures of local ferromagnetic order and of spin-dependent thermally-activated transport, which are described quantitatively by a simple phenomenological model. We discuss possible origins of the interfacial ferromagnetism. In addition, the *MR* also shows transient hysteretic features on a timescale of ~10-100 seconds. We demonstrate that these are consistent with an extrinsic magneto-thermal origin, which may have been misinterpreted in previous reports of magnetism in STO-based oxide interfaces. The existence of these two *MR* regimes (steady-state and transient) highlights the importance of time-dependent measurements for distinguishing signatures of ferromagnetism from other effects that can produce hysteresis at low temperatures.

The interface between two complex oxides can host a high-mobility conducting electron gas, even though the constituent materials are insulators. Since the initial discovery of this phenomenon in $LaAlO_3$/$SrTiO_3$ (LAO/STO) heterostructures [1], several other materials systems with this property have been identified [2-4]. These systems host a range of highly-correlated phases, including superconductivity [5,6] and ferromagnetism [7-12]. While superconductivity has been conclusively detected using transport [5,6,8,13] and scanning-probe measurements [9], reports of magnetism have been more difficult to reconcile across experiments. Numerous studies have investigated the magnetic properties of this class of oxide interfaces, with the majority of work to date focusing on the canonical LAO/STO system [7-10,14]. However, a unified picture of ferromagnetism is still lacking [7-9,15-18]. Magneto-transport studies [7,10,19], have reported pronounced magnetic hysteresis, suggestive of robust ferromagnetic order, however scanning-probe magnetometry [9] and neutron scattering [15] studies have generally detected low magnetic moments. One of the aims of the present work is to explore this inconsistency further.

A tantalizing possibility is that spin ordering emerges from electronic reconstruction at these interfaces [20,21]. However, given the variability across studies, it is also possible that the observed magnetic effects stem from sample-specific disorder, such as oxygen vacancies, dislocations and roughness [22-24]. Since LAO/STO is typically grown by pulsed-laser deposition (PLD) of LAO onto STO substrates, both stoichiometry and epitaxy near the interface have been difficult to control precisely [25]. Moreover, the STO substrate is typically an integral part of the interface. This poses an additional challenge since STO substrates contain defects which include unintentional magnetic impurities [26].

Some of these sources of disorder can be better controlled in related STO-based interface systems, which complement existing work on LAO/STO. Two promising but understudied systems are the titanate interfaces $GdTiO_3/SrTiO_3$ (GTO/STO) and $NdTiO_3/SrTiO_3$ (NTO/STO). These are grown using hybrid molecular beam epitaxy (hMBE), an ultra-high vacuum MBE technique recently developed for the growth of complex oxides [27], which ensures the highest possible control over stoichiometry at the interface [4,28,29]. Previous work on GTO/STO has shown evidence of interfacial ferromagnetism [10,14]. However, this could be linked to proximity-induced ferromagnetism, since GTO is ferrimagnetic and thus already has an intrinsic net moment.

Here we present the first investigation of the magnetic properties of NTO/STO heterostructures. Both constituent materials are grown via hMBE on an LSAT substrate, which, importantly, is not part of the interface. Furthermore, previous analysis of the transport and structural properties of NTO/STO indicates that oxygen vacancies, dislocations and substrate defects do not play a key role in its electronic properties [28,29], suggesting that any observed magnetic properties are likely not determined by these factors either. We report local ferromagnetism at the NTO/STO interface, and show that electronic transport is mediated by spin-dependent hopping between the localized ferromagnetic regions. We show that the magnetization of these regions fluctuates superparamagnetically and can become blocked at temperatures below ~1.5 K, giving rise to hysteresis. Finally, we also observe a distinct and more pronounced magnetic hysteresis. Crucially, however, our time-dependent measurements show that this effect is transient, calling into question previous reports of magnetic hysteresis in oxide interfaces. We demonstrate that this transient hysteresis, with characteristic time scales of 10-100 seconds, is in fact independent of the material being measured, and is consistent with heating of the sample due to extrinsic factors that are common in low-temperature measurements [30].

We investigate the magnetic properties of hMBE-grown NTO/STO using electronic transport measurements at temperatures down to 150 mK. Bulk NTO is a Mott-Hubbard anti-ferromagnetic insulator with Mott-Hubbard gap of 0.8 eV and Néel temperature of ~90 K [31], while STO is a band insulator with a gap of 3.2 eV. Electronic reconstruction at the NTO/STO interface forms a high-density quasi-two-dimensional electron gas (q2DEG), similar to the case of LAO/STO [32]. This q2DEG resides on the STO side of the NTO/STO interface, and extends for a few nm into the STO, as previously shown through band structure calculations based on x-ray photoemission spectroscopy (XPS) measurements and electron energy loss spectroscopy (EELS) [28]. The carrier concentration of stoichiometric NTO/STO heterostructures with less than 6 unit cells of NTO top layer is ~0.5 electrons per unit cell (u.c.) of STO ($3.2 \times 10^{14}$ cm$^{-2}$) [28], in excellent agreement with the prediction of the polar discontinuity model for polar-non-polar oxide interfaces [32]. Note that GTO/STO can also have this same charge density of ~0.5 electrons/u.c. [2]. In contrast, electron densities typically reported for LAO/STO samples are an order of magnitude lower than this predicted value. The exact reason for this discrepancy has not yet been resolved [29,33]. The charge density of ~0.5 electrons/u.c. in NTO/STO suggests that charged defects do not drive conduction in this system.

We focus on one sample (sample A) with layer thicknesses STO(40 u.c.)/NTO(4 u.c.)/(La$_{0.3}$Sr$_{0.7}$)(Al$_{0.65}$Ta$_{0.35}$)O$_3$ (LSAT) (001) (substrate). Data from four additional samples is provided in the Supplemental Materials. The temperature-dependence of the sheet resistance, $R_s$, is shown in Fig. 1(a) for different values of the magnetic field, $B$, applied in the plane of the sample. $R_s$ follows an Arrhenius-law behavior, $R_s \propto \exp[\varepsilon/k_B T]$, consistent with thermally-activated electronic transport, where $k_B$ is the Boltzmann constant, $T$ is the temperature and $\varepsilon$ is the activation energy. Remarkably, $\varepsilon$ decreases by over 60% of its $B=0$ value when a 6 T field is applied. This indicates that magnetic effects dominate transport through the sample at low temperatures. Such magnetically-tunable thermally-activated behavior is a robust characteristic of each of our STO/NTO samples with 4 u.c. NTO, which are the focus of this work. When the NTO thickness is increased to 10 u.c. the low-temperature transport shows lower resistance values and only a weak upturn in $R$ at low $T$ (Supplemental Fig. S8). This is consistent with the charge spillover mechanism present in NTO/STO interfaces with thicker NTO layers [28].

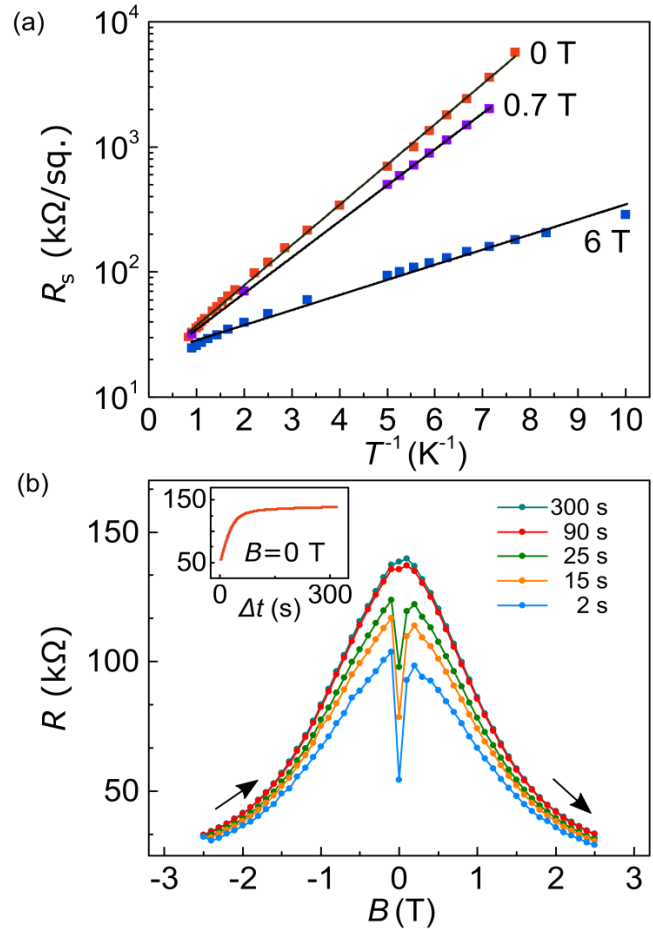

**Fig. 1: Sheet resistance and magnetoresistance.** (a) $R_s$ vs. $T^{-1}$ for three values of $B$. Lines are fits to $R_s \propto \exp[\varepsilon/k_B T]$. A weak deviation from a simple Arrhenius law occurs at larger values of $B$, indicating that $\varepsilon$ acquires a small $T$ - dependence as $B$ increases. (b) $R$ vs. $B$ at different values of $\Delta t$. Inset: $R$ vs. $\Delta t$ for $B$ = 0 T.

To further investigate the magnetic properties of the sample we measure the resistance, $R$, as a function of $B$, while also recording its time-dependence. Specifically, we step the value of $B$, then record $R$ as a function of $\Delta t$, the time elapsed since setting $B$ to its current value (a full two-dimensional plot of $R$ vs. $B$ and $\Delta t$ is shown in Supplemental Figure S1). Fig. 1(b) shows the magnetoresistance for several values of $\Delta t$. Two main features stand out. First, $R$ decreases by up to ~75% of its zero-field value upon application of a 2.5 T field. Second, curves measured at small $\Delta t$ exhibit a sharp resistance dip near $B$=0, which gradually decreases with increasing $\Delta t$ and eventually disappears for $\Delta t >$~90 s. The inset of Fig. 1(b) shows the time-evolution of *R* at a fixed value of *B*. Initially, *R* increases rapidly from a value of ~55 k$\Omega$, then saturates around a stable value of ~140 k$\Omega$, with a characteristic evolution time $\Delta t^* \sim 25 s$. We focus first on the steady-state properties ($\Delta t \gg \Delta t^*$), then return to the significance of the transient dip.

**Magneto-transport in the steady-state**

Important information about the underlying transport mechanism is obtained from the temperature evolution of the magneto-resistance, $MR \equiv \frac{[R(B)-R(0)]}{R(0)} \times 100\%$. At low temperatures, *R* becomes strongly suppressed for large *B*, yielding a large negative *MR* of up to -95% for *B* = 6 T at 150 mK (see Fig. 2(a)). As *T* is increased from 150 mK, the *MR* curves become gradually shallower and eventually acquire a slightly positive curvature for *T* >~4 K. In contrast to the large and strongly temperature-dependent negative *MR* observed below 4 K, above 4 K the *MR* does not exceed ~0.2% and is only weakly temperature-dependent. This suggests that distinct processes dominate transport above versus below 4 K. These observations are further supported by data from an additional thermally-activated sample (see Supplementary Fig. S7). Importantly, we also studied two samples where NTO was grown above STO (the opposite configuration from the samples discussed above) (see Figs. S5 and S6). These samples show small *MR* and qualitatively different *MR* curves. Ref.[28] shows through atomic-resolution STEM data that whenever STO is grown on NTO the interface is rough on the order of a few unit cells, while when NTO is grown on STO the interface is atomically-sharp, suggesting that interface roughness plays an important role in the observed *MR*.

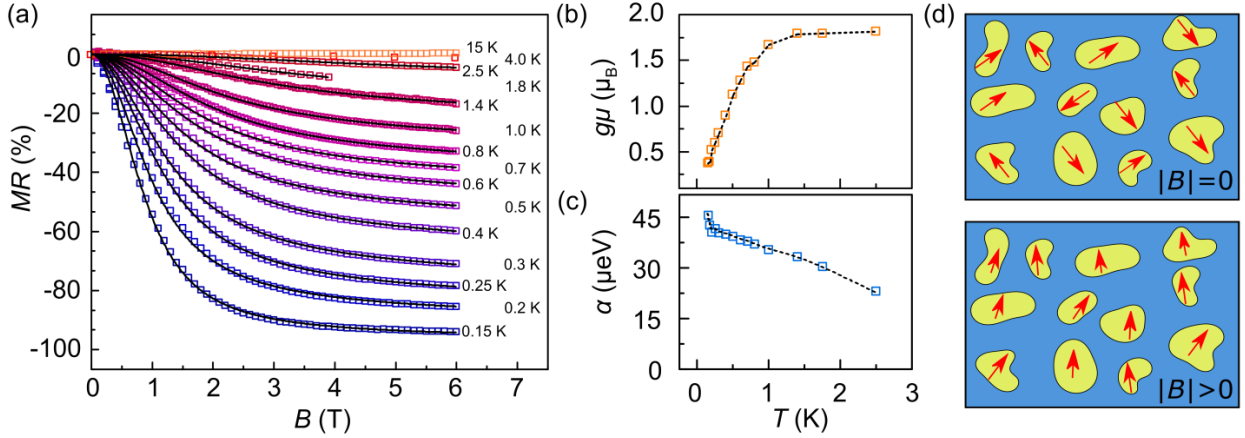

**Fig. 2: Magnetoresistance ratio and model fits.** (a) Magnetoresistance ratio at different values of $T$. Solid lines are fits to eqn. (1) for $T<\sim 4K$, where a pronounced negative $MR$ is observed. (b) Temperature evolution of the average $g\mu$. Values are obtained from the fits in (a). (c) Temperature evolution of the fitting parameter $\alpha$. The gradual decrease in $\alpha$ with increasing $T$ could be due to gradual loss of spin coherence during hopping at larger $T$. Dotted black lines in (b) & (c) are guides to the eye. (d) Schematic showing the net moments of the superparamagnetic islands discussed in the main text. Blue indicates non-ferromagnetic regions. At $B = 0$ the moments are randomly oriented (upper panel), while at finite $B$ they become polarized (lower panel).

Large negative $MR$ has been observed in LAO/STO [7,34,35], leading to several possible explanations, including ferromagnetic ordering [7], the Kondo effect [35,36], and a non-interacting model based on strong spin-orbit coupling [34]. Although these models capture certain features of the LAO/STO $MR$ data, a *quantitative* description of $MR$ curves in complex oxide interfaces at all values of $B$ is still missing. Below, we propose a simple model which accounts for all features of our data on NTO/STO samples with high accuracy. The model is intended for semi-insulating samples in which transport is mediated by hopping in the presence of an effective activation energy. The decrease in activation energy $\varepsilon$ with increasing $B$ (Fig. 1(a)) is reminiscent of carrier delocalization under applied magnetic field, reported in mixed-valence manganites which show colossal magnetoresistance [37,38]. Although the resemblance with manganites is purely phenomenological, it is useful to note that in those and other related materials [39] electrons are localized by a spin-dependent potential barrier, and become delocalized by an applied magnetic field. In our samples, the magnetic field dependence of $R$ is largely independent of the direction between $B$ and the sample plane (Supplemental Fig. S2), consistent with spin-dependent effects. The activation energy, $\varepsilon$, can be separated on general grounds into a spin-independent contribution, $\varepsilon_0$, and spin-dependent contribution, $\varepsilon_s$, yielding $\varepsilon = \varepsilon_0 + \varepsilon_s$. The spin-dependent energy cost for hopping from site 1 to site 2, with magnetization unit vectors $\boldsymbol{M}_1$ and $\boldsymbol{M}_2$ respectively, can be written as $\varepsilon_{12}^s = -\alpha \mathbf{M}_1 \cdot \mathbf{M}_2$, where $\alpha$ characterizes the hopping energy [38]. Since the observed MR is negative, $\alpha$ is positive, indicating that parallel alignment of spins decreases the energy cost. Macroscopic averaging leads to $\varepsilon_s = -\alpha (M/M_s)^2$, where $M$ is the magnetization and $M_s$ the saturation magnetization of the sample. In the paramagnetic limit, where $M$ vanishes for $B=0$, the MR can therefore be expressed as:

$$MR = \left( Exp\left[ -\frac{\alpha}{k_B T}\left(\frac{M}{M_s}\right)^2 \right] - 1 \right) \times 100\% \qquad (1)$$

For paramagnetic and superparamagnetic systems the local moments fluctuate and the average value of the $M$ is described by the Brillouin function, $B_j$, where $j$ is the angular momentum of each fluctuating moment. For larger values of $j$, fluctuating moments behave classically and the Brillouin function quickly approaches the Langevin function, yielding $\frac{\langle M \rangle}{M_s} = L(g\mu B/k_B T)$, where $g$ is the effective Landé g-factor and $\mu$ is the magnitude of the average fluctuating moment. To test the validity of our model, we fit equation (1) to the MR data (Fig. 2(a)). We find an excellent agreement over more than one order of magnitude in temperature using only two fitting parameters, $\alpha$ and $g\mu$. The extracted $\alpha$ parameter is comparable to the difference in activation energy at $B$=0T and $B$=6T shown in Fig. 1. This validates the applicability of the model. In particular, the strong temperature dependence and the large magnitude of the MR are consequences of the exponential factor in Eqn. (1). Interestingly, the Brillouin expression for the quantum paramagnet does not yield good fits for small $j < \sim 10\hbar$ when $T > \sim 1.5K$. This indicates that the angular momentum of each local moment is larger than $\sim 10\hbar$, which justifies using the classical description.

The remarkable fit of the Eqn. (1) model to the negative MR data points to spin-dependent hopping between magnetic regions for $T < 4$ K (Fig. 2(d)). In this picture, the moments of these regions fluctuate over time as described by the Langevin function. The lower bound on $j$ discussed above points to the magnetic regions being clusters with local ferromagnetic order, rather than individual atomic sites. Thus, $\mu$ is a measure of the average moment of the magnetic clusters that are fluctuating at a given temperature. Consistent with this picture, the values of $g\mu$ extracted from the fit are approximately constant for $T > 1.5$ K (Fig. 2(b)). This indicates that for 1.5 K $< T <$ 4 K all magnetic regions undergo large-amplitude superparamagnetic fluctuations. These magnetic particles may be completely free, but could also interact through exchange coupling, provided that the exchange interaction is weak compared to $k_BT$ [40]. On the other hand, for $T < 1.5$ K, $g\mu$ decreases rapidly and approaches zero near $T$=0. This suggests that below 1.5 K the regions with larger $\mu$ cease to fluctuate because $k_BT$ decreases below their magnetic anisotropy energy. Since the smaller regions have smaller anisotropy energy they continue to fluctuate as the temperature is decreased, until they are also gradually blocked at lower $T$. Thus, 1.5 K corresponds to an effective superparamagnetic blocking temperature of the entire sample. Note that the *MR* becomes weakly positive above 4 K, consistent with the dominance of orbital processes over spin effects above ~4 K (see Fig. 2(a) and Fig. S10). These observations are consistent with reports of weak local ferromagnetism and superparamagnetism observed by scanning SQUID studies in other STO-based interfaces[9,12,24].

Below ~800 mK the *MR* curves become hysteretic (Fig. 3), consistent with a large fraction of the superparamagnetic domains becoming cooled below their blocking temperature. In this temperature range, magnetic anisotropy energies dominate over thermal effects. This stabilizes the direction of the local domains and leads to irreversible behavior. Consistent with this picture, the hysteresis becomes more pronounced with decreasing temperature (Fig. 3). The presence of hysteresis provides direct evidence for ferromagnetic ordering. We note that hysteresis is a sufficient, but not necessary signature of local ferromagnetism. Its presence only indicates that at low $T$ the moments of the magnetic clusters do not fluctuate over the duration of the measurement.

Although a full microscopic description of the NTO/STO interface is beyond the scope of this study, the small extracted $g\mu$ values could be indicative of local canting of spins at the interface and with the presence of so-called weak ferromagnetism. Weak ferromagnetism is a form of ferromagnetism that occurs in certain antiferromagnets, in which the Dzyaloshinskii-Moriya interaction stabilizes spin canting, leading to a small spontaneous net moment. The Dzyaloshinskii-Moriya interaction is allowed in rare-earth titanates, and past work on doped $NdTiO_3$ showed that a weak ferromagnetic phase can form in this material due to spin canting, with measured net moment densities of ~$10^{-3}$ to $10^{-2}$ $\mu_B$/formula unit, depending on doping level [31]. In our nominally undoped samples we can expect that weak antiferromagnetism, if present, could be further affected by the ultra-high interfacial charge density and by strong local electric fields stemming from interface roughness (see the Supplementary Materials for further discussion of roughness in our samples). Moreover, we note that our fit yields the product $g\mu$, rather than just the moment $\mu$. Past work has shown that $g$ itself can be quite small in STO-based interfacial electron systems, with values as low as ~0.6 [41]. This shows that the small extracted $g\mu$ values are consistent with weak ferromagnetism in our samples.

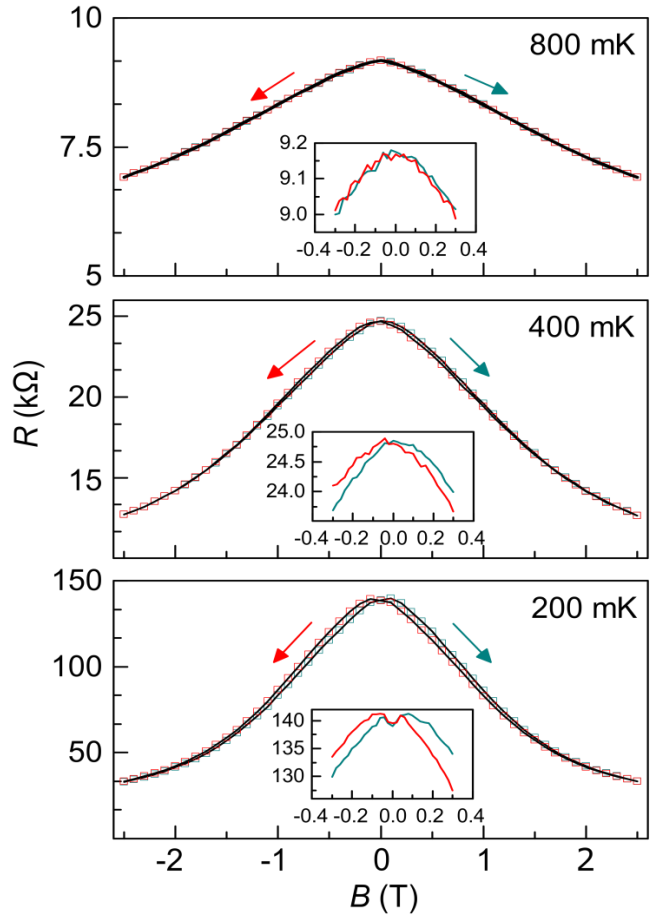

**Fig. 3: Magnetoresistance hysteresis.** $R$ vs. $B$ while stepping $B$ from negative to positive value (cyan) and positive to negative values (red). Black lines are guides to the eye. Arrows indicate the direction along which $B$ is stepped. Insets in each panel show a zoom-in near $B$=0, highlighting that hysteresis becomes more pronounced as $T$ is lowered.

**Magneto-transport in the transient regime**

Unlike the data presented in the previous section, conventional *MR* measurements (including continuous sweep methods) are performed effectively in the limit of small $\Delta t$. As noted above and shown in Fig. 1(b), in this time regime, $\Delta t < \Delta t^*$, we also observe magnetic hysteresis. In fact, for $\Delta t < \Delta t^*$ the hysteresis is far more pronounced than that observed in the steady-state. This is illustrated in Fig. 4(a), which shows the *MR* measured at $\Delta t = 2\,\text{s}$ and *T*=200 mK (bottom panel). The data shows a striking hysteretic dip for |*B*|< ~50 mT. Such butterfly-shaped dips in the magneto-resistance have been previously used as evidence of interfacial magnetism in oxide interfaces and oxide thin films [7,10,19]. However, given the transient nature of the *MR* observed here (as well as in Ref.[7]), it is essential to establish whether this effect is related to magnetism in the sample or if it has a different origin. To investigate its origin, we performed a control experiment on a non-magnetic, Si-doped, insulating GaAs crystal with similar resistance to that of sample A (Fig. 4(b)). Si-GaAs is an excellent control sample because it is one of the best understood semiconductors. Strikingly, the *MR* curve measured on GaAs at $\Delta t = 2\,\text{s}$ shows a similar hysteretic butterfly-shaped dip (Fig. 4(b), bottom panel) as the NTO/STO sample. This feature is also transient, as shown in Fig. 4(b), upper panel, measured for $\Delta t = 320\,\text{s}$. Since the GaAs sample is a high-purity non-magnetic bulk crystal, this demonstrates that this transient hysteretic resistance dip occurs independently of the system used and is therefore not the result of magnetism in the sample, but is instead extrinsic to the sample. This conclusion is further supported by a similar measurement on a BaSnO$_3$ thin film (Fig. S4).

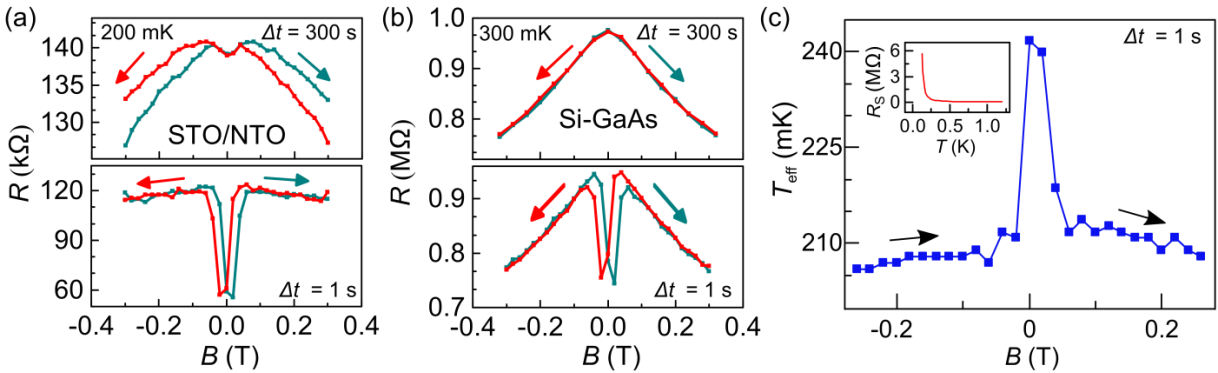

**Fig. 4: Magnetoresistance at short $\Delta t$.** *R* vs. *B* for: (a) sample A and (b) a Si-GaAs control sample. For both samples, a hysteretic dip develops near *B* = 0 at short $\Delta t$ (lower panels), while no such dip is observed at long $\Delta t$ (upper panels). Unlike the NTO/STO sample, the GaAs sample does not show any hysteresis in the steady state, but only a weak, reversible, negative *MR*, consistent with quantum interference effects in insulating n-GaAs [42]. (c) Estimated effective temperature of sample A as a function of *B* for $\Delta t = 2\,\text{s}$. Arrows indicate the direction along which *B* is stepped. Inset: *R* vs. *T* for sample A, showing the steep dependence of *R* with *T* at low values of *T*, characteristic of semi-insulating behavior.

Having ruled out intrinsic magnetism as the origin of the transient hysteresis, we now consider the possibility of a magneto-thermal origin extrinsic to the sample. Interestingly, of all the samples we studied, the transient dip was observed only in samples for which $R$ has a large upturn at low $T$ (e.g. Fig. 1(a) and inset of Fig. 4(c)). In contrast, metallic samples, for which $R$ is only weakly temperature dependent at low $T$, show no transient hysteresis. This is consistent with a magneto-thermal origin of the transient hysteresis dip, which is extrinsic to the sample itself: for semi-insulating samples even a small increase in the effective sample temperature leads to a substantial decrease in $R$.

To quantify this effect, we estimated the effective temperature of our STO/NTO/LSAT sample as a function of $B$ at $\Delta t = 2$s (Fig. 4(c)) (see Supplemental Figure S3 for further details). We find that the transient hysteretic effect is associated with an increase in sample temperature by only ~30 mK, which develops as $B$ is stepped through the -80 mT to +80 mT range. This rapid increase in temperature near $B=0$ could be due to heat released by an avalanche magnetic reversal process. Such processes are commonly observed when the spin configuration of a ferromagnet, spin glass or spin-ice switches abruptly [43]. The presence of a magneto-thermal material near our sample is confirmed by the fact that the mixing chamber temperature of our dilution refrigerator rises by ~5 mK as $B$ crosses zero. While we cannot precisely determine the location of the magneto-thermal material, we note that magneto-thermal effects are common and can have dramatic consequences [30] below several Kelvin, where the relative change in $T$ can be large. Both the hysteretic and the transient nature of the butterfly-shaped resistance dip can be understood as consequences of the characteristic thermal re-equilibration time of the sample, which in our case is only $\sim \Delta t^* \sim 25$s (Fig. 1(b) and Fig. S1). The key point is that magneto-thermal heating (originating outside the sample) changes the temperature (and hence the resistance) of the sample and that $R$ then takes a finite time to recover as the heat is dissipated. This finite thermal re-equilibration time leads to the observed hysteresis.

It is important to note that in the absence of time-dependent data such as this (which has generally not been investigated in complex oxide studies) the intrinsic magnetic hysteresis in NTO/STO would have been obscured by this much larger magneto-thermal hysteresis, and its transient, extrinsic nature would have been hard to detect. This *MR* effect is universal and applies to any material with a sufficiently large temperature-dependence of the resistance at low $T$, including superconductors near their critical temperatures and upper critical fields. It is also expected to be particularly effective in known magneto-caloric materials such as the ferromagnetic alloys of Gd [44]. Note that the possibility of a magneto-thermal origin for the *MR* hysteresis reported in LAO/STO in Ref. [7] has also been discussed in Ref. [45].

In conclusion, we observe weak local ferromagnetic ordering at the epitaxial interface between NTO and STO. Our analysis suggests that ferromagnetism develops inhomogeneously in isolated regions, which are superparamagnetic down to very low temperatures, but can become blocked below ~1.5 K. Electronic transport is in excellent quantitative agreement with spin-dependent, thermally-activated hopping between the ferromagnetic regions. Oxygen vacancies, dislocation defects and substrate impurities are unlikely to play a key role in the observed magnetism since these factors have no detectable role on the electronic properties of our NTO/STO heterostructures, as shown using STEM-EDX, STEM-EELS, XPS and transport data in Refs. [29] and [28]. Comparison with samples with smoother interfaces suggests that interfacial roughness likely plays an important role, possibly by generating regions of uncompensated spins in antiferromagnetic NTO at the interface. This is supported by the fact that large negative *MR* ratios are observed only in heterostructures where STO was grown on NTO, but not in samples where the growth order was reversed. The layer order is significant because atomic resolution electron microscopy shows STO/NTO interfaces to be rough on the atomic scale due to intermixing, while the inverted order (NTO/STO) results in atomically sharp interfaces [28]. The Dzyaloshinskii-Moriya interaction could also contribute to the onset of the observed ferromagnetism in conjunction with the interfacial roughness and the large interfacial charge densities [31]. Importantly, we also highlight the need for time-dependent measurements for distinguishing signatures of magnetic order from commonly-occurring extrinsic magneto-thermal effects that can dramatically affect the interpretation of low-temperature data on a wide range of materials.

**Acknowledgements:**

The authors thank P. Crowell, E. D. Dahlberg, J. Levy, A. Morpurgo, R. Fernandes and B. Shklovskii for valuable discussions. This work was supported primarily by the National Science Foundation (NSF) Materials Research Science and Engineering Center (MRSEC) at the University of Minnesota Under Award Number DMR-1420013, and through the Young Investigator Program of the Air Force Office of Scientific Research (AFOSR) through Grant FA9550-16-1-0205. Device fabrication was performed in part at the Minnesota Nano Center (MNC), which is partially supported by the NSF through the National Nanotechnology Coordinated Infrastructure (NNCI).

# Ferromagnetism and spin-dependent transport at a complex oxide interface

Yilikal Ayino[1], Peng Xu[2], Juan Tigre-Lazo[1], Jin Yue[2], Bharat Jalan[2], and Vlad S. Pribiag[1*]

1. School of Physics and Astronomy, University of Minnesota, Minneapolis, MN 55455, USA

2. Department of Chemical Engineering and Materials Science, University of Minnesota, Minneapolis, MN 55455, USA

*e-mail: vpribiag@umn.edu

**List of supplemental figures and discussions:**

**Additional details regarding the measurements**

**Procedure for the time-dependent magnetoresistance measurements**

**Angular dependence of magnetoresistance**

**Supplemental Fig. S1:** Time-dependent magnetoresistance data

**Supplemental Fig. S2:** Angular dependence of resistance

**Estimation of sample temperature due to extrinsic magneto-thermal heating**
    **Supplemental Fig. S3:** Estimation of sample temperature due to extrinsic magneto-thermal heating

**Magnetoresistance of a $BaSnO_3$ control sample**
    **Supplemental Fig. S4:** Magnetoresistance of a $BaSnO_3$ control sample.

**NTO/STO/LSAT samples (inverted layer sequence with respect to sample A in main text)**
    **Supplemental Fig. S5:** Data for sample D, a NTO (20u.c.)/STO (24u.c.)/LSAT sample (reversed growth sequence with respect to samples A-C).
    **Supplemental Fig. S6:** Data for sample E, a NTO (10u.c.)/STO (50 u.c.)/LSAT sample (reversed growth sequence with respect to samples A-C).

**Transport data from additional non-inverted samples**
    **Fig. S7:** Data from sample B (STO(8uc)/NTO(4uc)/LSAT).
    **Fig. S8:** Data from sample C (STO(24uc)/NTO(10uc)/LSAT).

**Supplemental Fig. S9:** Voltage vs. current curves vs. *T* for sample A.

**Other models of large negative *MR***

***MR* data on sample A between 5K and 10K**
    **Fig. S10:** MR data on sample A above 4 K

**Quantitative analysis of the NTO/STO interface**

**Discussion of the spin-dependent hopping model and its comparison with microscopic phase separation in mixed-valence manganites**

**References**

**Additional details regarding the measurements**

Measurements were performed in a four-terminal van der Pauw configuration, in a dilution refrigerator equipped with three-axis vector rotate magnet, with a base temperature of 15 mK. Due to the thermally-activated nature of transport in our samples, the current-voltage (*I-V*) characteristics become nonlinear for T <~150 mK, indicating non-Ohmic transport (see Figure S9). We therefore focused on $T \geq$ 150 mK. Samples were contacted by wire-bonding to sputtered Al (40nm)/Ni(20nm)/Au(300nm) contacts. We studied in detail three samples with layer structures STO(*t* u.c.)/NTO(*n* u.c.)/LSAT, where *t* was varied between 8 and 40 u.c. and *n* was 4 or 10 u.c. The data from the additional samples are shown below. Persistent hysteresis was observed only in sample A, discussed in the main text. The lack of detectable hysteresis for $\Delta t \gg \Delta t^*$ in the other samples indicates very low magnetic coercivities in those samples (see Fig. S7 for a further discussion of this point). This further suggests that the irreversible contributions to ferromagnetic reversal are sensitive to sample-specific disorder, likely interfacial roughness. This is point is elaborated below, through an explicit comparison with two inverted samples, NTO(*n* u.c.)/STO(*t* u.c.)/LSAT (see Fig. S5 and Fig. S6).

**Procedure for the time-dependent magnetoresistance measurements:**

The time-dependent *MR* is measured using the following procedure. Before starting the measurement *B* is first set at some large negative or positive value. During the measurement, *B* is stepped and held constant while the resistance is measured as a function of time delay $\Delta t$ under constant B. The field is stepped again after when $\Delta t$ reaches 320 s, then the procedure is repeated. All measurements are performed under constant current bias. This procedure generates a three-dimensional data set: *R* as a function of *B* and $\Delta t$ (Fig. S1). This is represented in the 2D color plot with magnetic field and delay in each axis and resistance represented by a color scheme, Fig S1 (a). It can also be represented in 1D plot of resistance as a function of the total time elapsed since the beginning of the measurement S1 (b).

The timescale for the suppression of the butterfly-shaped hysteretic dip (Fig. 1b and Supplemental Figure S1) suggests a characteristic timescale for temperature re-equilibration of ~$\Delta t^* \sim 25$ s. This highlights the importance of analyzing the time-dependence of the *MR* when investigating signatures of magnetism in oxide interfaces or other materials for which the resistance is strongly temperature-dependent at low temperatures.

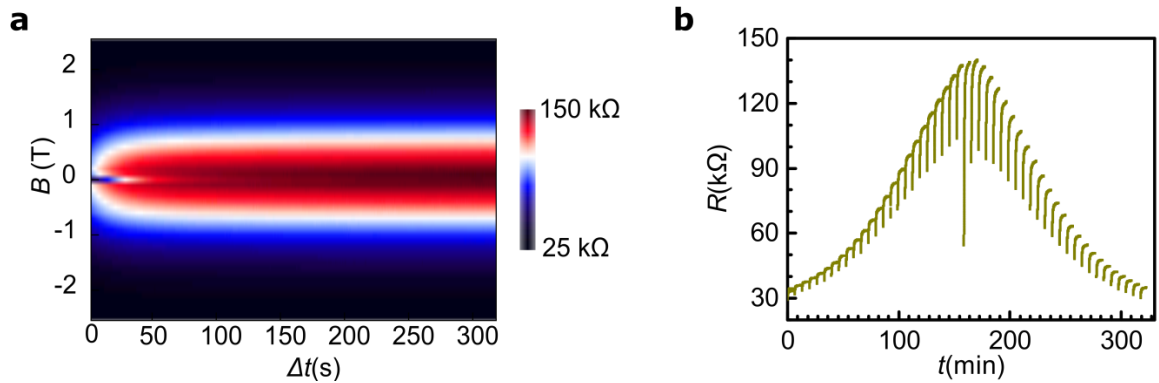

**Fig. S1: Time-dependent magnetoresistance data. a**, Two-dimensional color plot of R as a function of *B* and $\Delta t$. **b**, One-dimensional plot of the time-evolution of *R* plotted vs. the total time elapsed from the beginning of the measurement. Each curve represents resistance measured at a *B*. *B* is stepped during the time intervals which separate the curves, at which point the resistance is not measured.

**Angular dependence of magnetoresistance:**

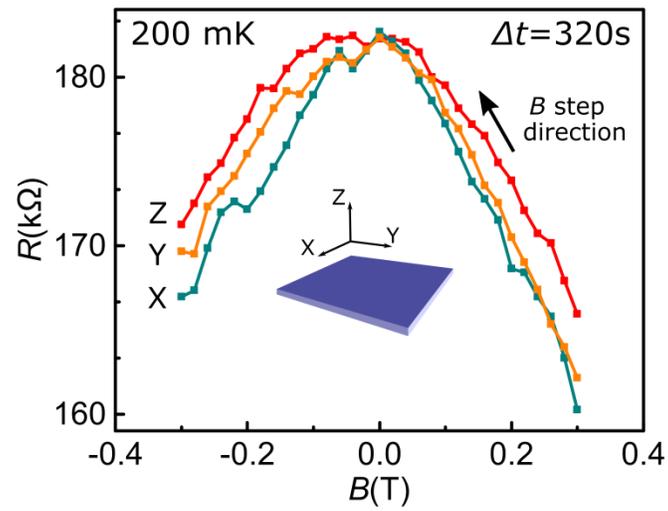

**Fig. S2: Angular dependence of resistance.** The resistance is largely independent of the direction of the applied field *B* with respect to the sample plane. The *MR* is slightly smaller when *B* is applied out of plane (along Z) than in the plane (along X or Y). This is consistent with a weak in-plane magnetic anisotropy.

**Estimation of sample temperature due to extrinsic magneto-thermal heating:**

As discussed in the main text, the resistance of the sample is a strong function of $T$, $B$ and $\Delta t$. In order to determine the effective sample temperature ($T_{eff}$) at a given $B$ and $\Delta t$ we use the steady-state $MR$ curves taken at $\Delta t \gg \Delta t^*$ as reference. Specifically, we assume that for $\Delta t \gg \Delta t^*$, $T_{eff} = T_{mix}$, where $T_{mix}$ is the mixing chamber temperature. To acquire the calibration data, we measure $R$ as a function of $B$ and $\Delta t$ for $T_{mix}$ ranging from 200 mK to 270 mK in 5 mK increments. At each temperature, we ensure that $T_{mix}$ has stabilized near the desired setpoint before measuring $R$. The resulting steady-state data $R(\Delta t=320$ s, $T_{mix}, B)$ are shown in Fig. S2.

To determine $T_{eff}$ as a function of $B$ in the transient regime ($\Delta t < \Delta t^*$) when $T_{mix}$ is set to 200 mK, we compare the measured values of $R(\Delta t, T_{mix}=200$ mK, $B)$ with the steady-state calibration data, $R(\Delta t=320$ s, $T_{mix}, B)$. This provides an estimate of $T_{eff}$ at short delay for $T_{mix}$ =200mK at each value of B. Because of the finite resolution of $T_{mix}$ increments (5 mK), we use spline interpolation to estimate temperatures from resistance values which fall in between adjacent points. Hence the uncertainty of $T_{eff}$ is estimated to be +/- 2.5mK.

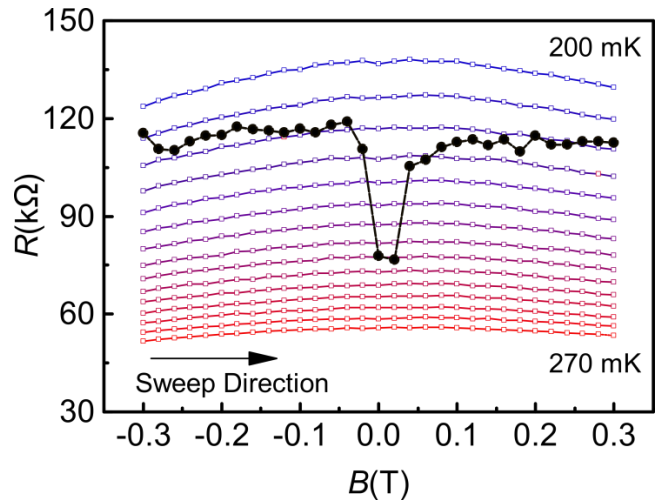

**Fig. S3: Estimation of sample temperature due to extrinsic magneto-thermal heating.** Curves with open points represent long delay resistance, $R(\Delta t=320$ s, $T_{mix}, B)$, for $T_{mix}$ ranging from 200 mK to 270 mK in 5 mK increments. The overlaid curve with solid points is the short delay data, $R(\Delta t = 2s, T_{mix}=200$ mK, $B)$, showing that near zero field a temperature spike of approximately 30-40 mK occurs for $T_{mix}$ = 200 mK.

**Magnetoresistance of a BaSnO₃ control sample:**

In addition to the Si-GaAs control sample discussed in Fig. 4 of the main text, we also performed a similar control experiment on a semi-insulating BaSnO₃ thin film. We found the same behavior: a hysteretic butterfly-shaped dip appears at low $\Delta t$ and vanishes in the steady-state, for $\Delta t > \sim 200$ s (Fig. S4). The only samples which did not show a transient butterfly-shaped dip were metallic (NTO/STO) samples for which the resistance is a weak function of temperature at low temperatures. This is consistent with a heating origin of the resistance dip, as discussed in the main text.

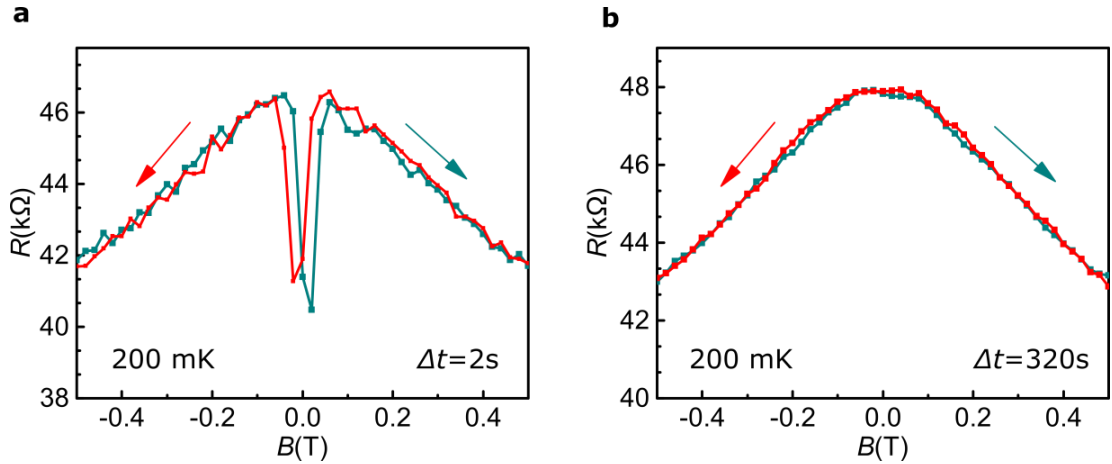

**Fig. S4: Magnetoresistance of a BaSnO₃ control sample. a**, $R$ as a function of $B$ at short delay ($\Delta t = 2s$) for control sample thin film consisting of a La-doped BaSnO₃ (BSO) thin film (layer structure is La-BSO(100nm)/BSO(120 nm)/STO (substrate)). The resistance shows a hysteretic dip near zero field, similar to that seen for STO/NTO/LSAT samples, as discussed in the main text. **b**, Resistance as function of magnetic field at ($\Delta t = 320s$), showing that the dip disappears in the steady state.

**NTO/STO/LSAT samples (inverted layer sequence with respect to sample A in main text):**

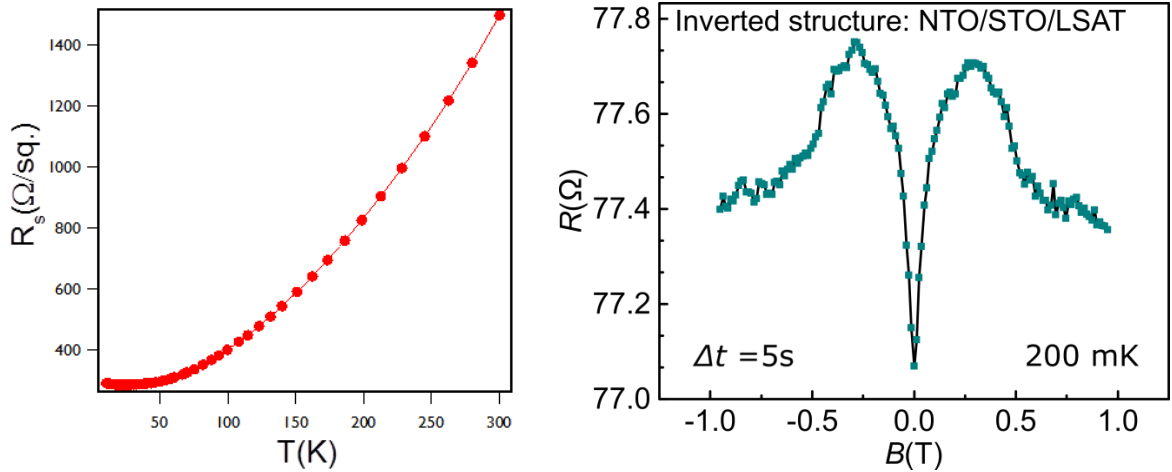

**Fig. S5: Data for sample D, a NTO/STO/LSAT sample (reversed growth sequence with respect to samples A-C).** The left panel shows the temperature-dependence of the sheet resistance. The right panel shows *R* vs. *B* for this NTO(20 u.c.)/STO(24 u.c)/LSAT heterostructure. Note that the NTO layer grown on the STO layer, which is opposite to the order for sample A, which is discussed in the main text (data are for $\Delta t \gg \Delta t^*$). The data shows a small sharp positive *MR* near B=0. Away from $|B| > \sim$0.3 T there is a small decrease in resistance. The absence of large negative *MR* is characteristic of heterostructures grown with such "inverted" layer ordering, which have an atomically-sharp interface (see also data on inverted sample E, below). Note the small negative *MR* (max *MR* ~0.9 %) and the qualitatively-different shape with respect to non-inverted samples (samples A-C). This is likely due to the much lower interface roughness in inverted samples (see STEM images in Ref. S1). Note also that this sample has the same NTO thickness as the non-inverted sample C (below).

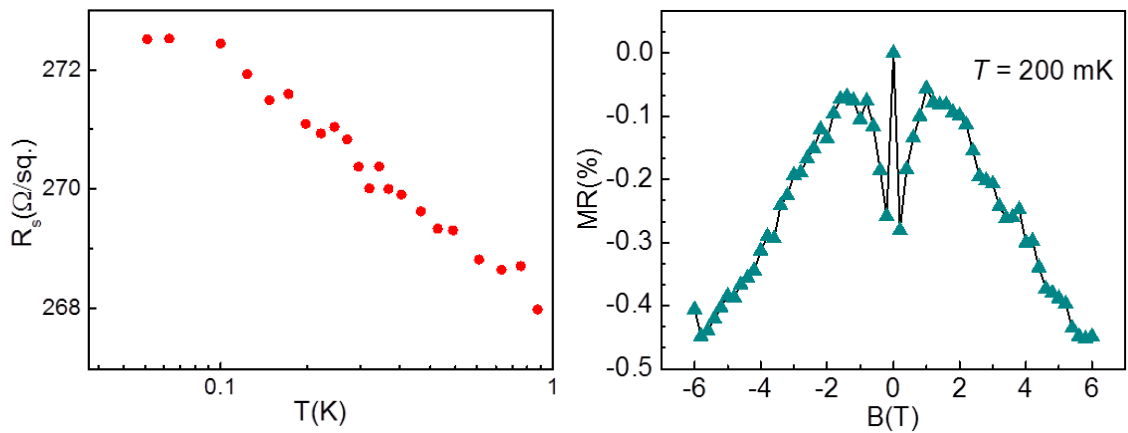

**Fig. S6: Data from sample E (inverted: NTO(10uc)/STO(50uc)/LSAT).** The left panel shows the temperature-dependence of the sheet resistance, illustrating that transport is not described by an Arrhenius law. The right panel shows that the sample has a small negative *MR* (max MR ~-0.5%) (data are for $\Delta t \gg \Delta t^*$). In this sample, the *MR* is also qualitatively different from that in non-inverted

samples (A-C), but, importantly, it is largely similar to that of the other inverted sample (sample D). Again, this is likely due to the much lower interface roughness in inverted samples.

**Transport data from additional non-inverted samples, with thin and thick NTO:**

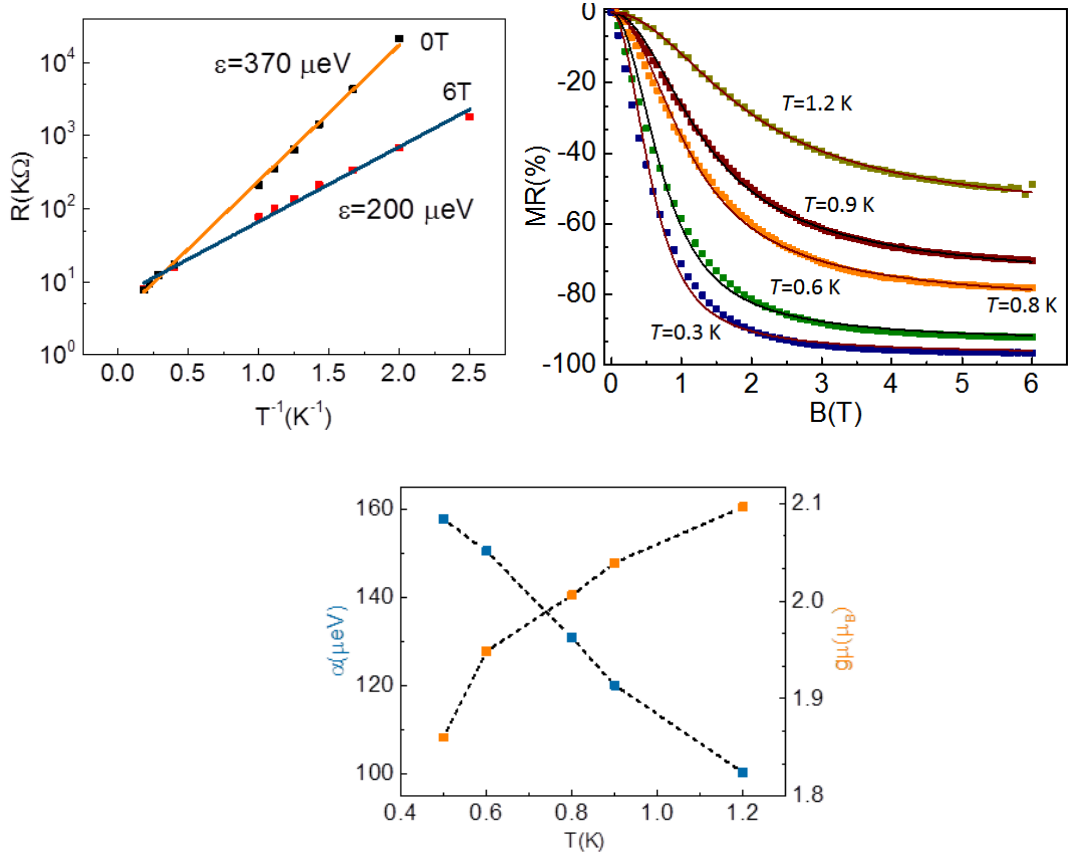

**Fig. S7: Data from sample B (STO(8uc)/NTO(4uc)/LSAT).** The top left panel shows that transport is thermally activated. Lines correspond to fits to $\exp[\frac{\varepsilon}{k_B T}]$. Evident from the plot is the decrease in activation under applied magnetic field, similar to sample A. The top right panel shows that the sample has large negative *MR*. Lines are fits to our model of spin-dependent scattering (eqn. 1 in the main text) (data are for $\Delta t \gg \Delta t^*$). The maximum *MR* is ~ -95%, comparable to the other semi-insulating sample (sample A, discussed in the main text). The lower plot shows the fit parameters. Like in sample A, the extracted $\alpha$ parameter is comparable to the difference in activation energy at 0T and 6T. This is indicative of the validity of applying the model. Note that unlike in sample A (main text) $g\mu$ for sample B does not approach zero as the temperature is lowered. This indicates that even magnetic regions with larger $\mu$ continue to fluctuate down to the lowest temperatures. This suggests that the magnetic anisotropies are very small in sample B, consistent with the absence of hysteresis, even though local ferromagnetism is present.

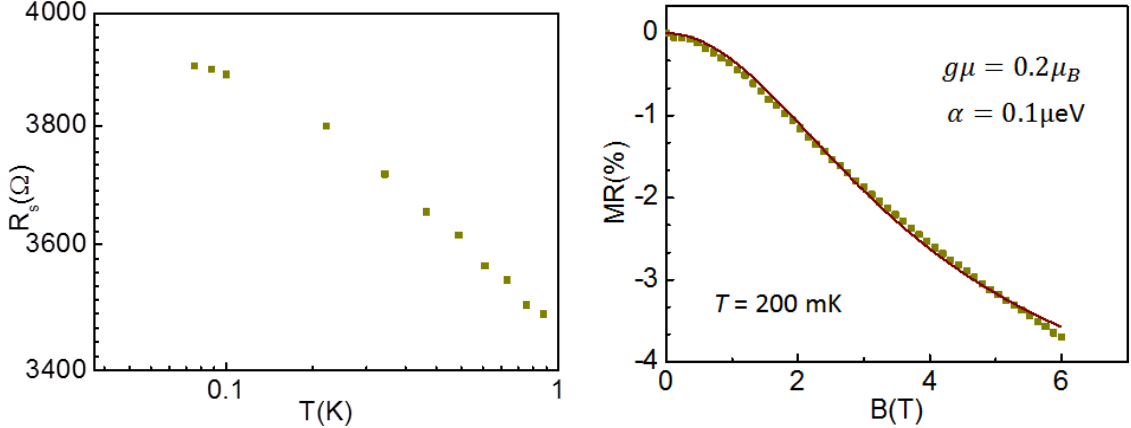

**Fig. S8: Data from sample C (STO(24uc)/NTO(10uc)/LSAT).** The left panel shows the temperature-dependence of the sheet resistance, illustrating that transport is not described by an Arrhenius law. The right panel shows that the sample has a small negative *MR* (max *MR* only -4%), consistent with our model, which predicts that large *MR* is associated with thermally activated-transport in our NTO/STO samples (data are for $\Delta t \gg \Delta t^*$). Instead, for sample C, *R* shows only a weak upturn at low *T*. Correspondingly, the fit of the *MR* data to our spin-dependent transport model (red line) is not as good as for the thermally-activated samples (Samples A and B), and yields an effective activation energy parameter $\alpha$ of only 0.1 μeV, about three orders of magnitude lower than for the thermally-activated samples, further validating the applicability and sensitivity of our model. The extracted value of $\alpha = 0.1\mu eV$ is two orders of magnitude smaller than the temperature scale of 0.2K (~ 17 μeV), consistent with non-Arrhenius behavior of transport. This implies that magneto-transport in this sample is not due to tuning of a spin-dependent activation energy. Note that, in contrast, for samples A and B the extracted $\alpha$ parameter is comparable or bigger than the temperature scale. This serves as a consistency check on the model: large *MR* occurs if and only if the sample shows thermally-activated transport.

To summarize, the two semi-insulating STO/NTO samples (Samples A and B, 4 u.c. NTO) both showed a large negative *MR* (~-95%). Both the large *MR* and the shapes of the *MR* curves for these samples are in quantitative agreement with our model of spin-dependent transport. In addition, we studied a third sample with 10 u.c. NTO (sample C), which does not display thermally-activated transport and has much lower resistances. The lower *R* at low *T* in Sample C is understood in terms of the charge transfer mechanism described by us in Ref. S1, which affects NTO/STO samples with thicker NTO and larger strain. This sample has a much weaker negative *MR* of only -4%, and shows a worse fit to our model than samples A and B. Importantly, our model yields an effective activation energy parameter $\alpha$ that is about three orders of magnitude smaller than for the thermally-activated samples (Samples A and B). This serves as a consistency check on the model: large *MR* occurs if and only if the sample shows thermally-activated transport.

Furthermore, we studied two inverted samples (NTO/STO), one with 20 u.c. NTO (sample D, Fig. S5) and one with 10 u.c. NTO (sample E, Fig. S6). Both inverted samples show very weak *MR* (less than 1%). Importantly, the shape of the *MR* is qualitatively different from that observed in the non-inverted samples near B=0. This suggests that the difference in roughness between non-inverted and inverted interfaces (see STEM microscopy images in Ref. S1) plays an important role for the observed magnetism, as discussed in the manuscript.

**Comparison of the *MR* of an STO/NTO/LSAT sample with an inverted (NTO/STO/LSAT) sample, both with the same NTO thickness:**

By comparing the data in Fig. S8 (STO/NTO/LSAT sample C) to those in Fig. S6 (inverted, NTO/STO/LSAT sample E), we show that the *MR* is qualitatively different for the two cases, even when both have the same NTO thickness. The *MR* of the inverted sample (Sample C) is monotonically negative, while that of the inverted sample (Sample E) shows a complex, non-monotonic structure, with a large dip and peak near B=0. While the origin of this structure in the inverted samples (a similar structure, minus the peak is also seen in inverted sample D) remains an open question, it could be due to with weak (anti-)localization. Note also that the MR is several times smaller in the inverted sample E than in the non-inverted sample C, another qualitative difference.

As stated in the main text, and evidenced via Scanning Transmission Electron Microscopy (STEM) in Ref. S1, the non-inverted interface (STO/NTO) is rough, while the inverted interface (NTO/STO) is atomically sharp. This dramatic difference in the smoothness of the two types of interfaces, together with the qualitative difference in *MR* discussed above, suggests that interfacial roughness likely plays an important role for the observed inhomogeneous magnetic phase, though we cannot exclude the possibility that other mechanisms may also be relevant. Roughness could induce local ferromagnetic order at the non-inverted (STO above NTO) interface by generating uncompensated spins. One possible mechanism is the formation of so-called weak ferromagnetism due to the Dzyaloshinskii-Moriya interaction in conjunction with the ultra-high interface densities and with the large local electric fields expected from interfacial roughness (see main text for references). This candidate microscopic picture is in accord with the small value of the magnetic moment that is suggested by the extracted fitting parameters of our model (assuming a g-factor of ~2) (see Fig. 2b and S7).

A second, but related possibility could be that the observed interfacial roughness in STO/NTO may lead to some interfacial intermixing. Specifically, this could create small clusters of Nd and $Ti^{3+}$ atoms (which are spinful) in the STO near the interface (see the quantitative analysis of the interface in Ref. [S1]).

Note that previous work on $SmTiO_3$/$SrTiO_3$, where $SmTiO_3$ is also antiferromagnetic, did not show evidence of interfacial ferromagnetism.[S2] This further supports the picture that the specific details of the interface, such as roughness, play an important role for the observed magnetism in NTO/STO.

**Voltage vs. current curves at low temperatures:**

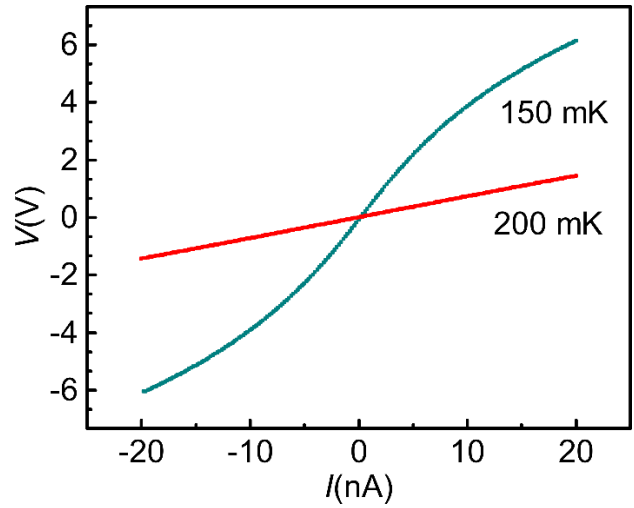

**Fig. S9: Voltage vs. current curves vs. *T* for sample A.** Voltage drop (*V*) across the sample as a function of applied bias current (*I*) for *T*=150 mK and *T*=250 mK. For *T* = 150 mK the *V-I* curve becomes nonlinear for *I* > ~3nA. For this reason, the data in the main text was taken at *T*≥150 mK.

**Other models of large negative *MR*:**

Previous work has shown that large negative *MR* in oxide interfaces can be obtained due to Kondo physics or due to large spin-orbit coupling in LAO/STO interfaces. Below, we show that these models do not account for the large negative *MR* observed in our semi-insulating STO/NTO samples.

Kondo model: The Kondo model (developed theoretically in Refs. S3 and S4) predicts large *MR* only for in-plane *B*-field. In contrast, our devices show large *MR* for both in-plane and out-of-plane *B*, with no substantial anisotropy (Fig. S2). Second, the Kondo model does not apply to our semi-insulating samples, which show exponential temperature-dependence and are therefore not in the Kondo regime. (For the Kondo regime, a logarithmic temperature-dependence is expected at low temperatures.) Third, the Kondo model does not account for the observed hysteresis. The steady-state *MR* hysteresis can only originate from irreversible reorientation of ferromagnetic domains, which is in agreement with our model.

Non-interacting model with strong spin-orbital coupling: The model based on strong spin-orbit coupling (Ref. S5) is by construction limited to *MR* of at most -50%. However, our data shows *MR* in excess of -90%. This model also applies only to in-plane fields, however our data (Fig. S2) shows that the effect is also present for out-of-plane field. We therefore conclude that this model does not apply to our data.

***MR* data on sample A between 5K and 100K:**

Our model (eqn. 1 of the main text) shows that negative *MR* in our samples is due to local ferromagnetism. Our data (Fig. 2a) shows negative *MR* for $T \leq 4$ K and positive *MR* above 4K, which therefore sets a lower bound on $T_c$ of ~4 K. Note that since the ferromagnetic regions are already fluctuating superparamagneticaly at 4K, it is fundamentally impossible to set an upper bound on $T_c$ using time-average data (i.e. any typical transport or magnetometry measurement).

The data on sample A (Fig. 2a) show that 4K marks a transition from negative to weakly positive *MR*. The data below track the *MR* from 5K to 100 K. At 5K and above the *MR* is weakly positive, indicating that weak orbital effects dominate over any remaining spin effects for $T > 4$K. As *T* is increased further from 10K to 100 K the *MR* becomes independent of *B*, indicating that by 100 K the orbital effects are smeared out by temperature.

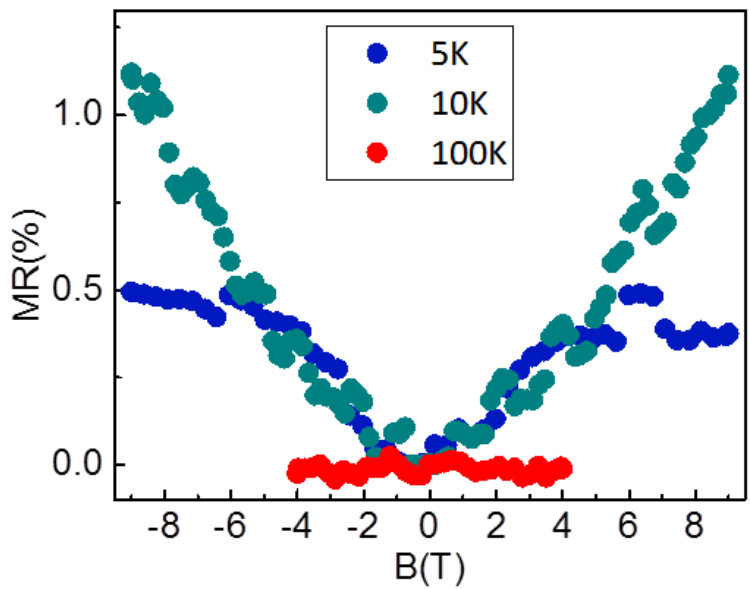

**Fig. S10: Magnetoresistance of sample A from for *T* ranging from 5 K to 100 K.**

**Quantitative analysis of the NTO/STO interface:**

Using scanning transmission electron microscopy with electron energy loss spectroscopy (STEM-EELS) and energy dispersive x-ray spectroscopy (EDX) mapping, we previously showed that cation intermixing/oxygen vacancies/non-stoichiometry cannot explain the observed $n_{2d}$ in our films [S1]. In Ref. S1, we used HAADF intensity profile (Fig. 2a in Ref. S1) and the elemental maps (Fig. S4b-h in Ref. S1) to show the normalized O, Sr and Nd profiles across the NTO/STO interface. The oxygen signal was normalized to the substrate value, and the Sr and Nd signals were normalized to the value within the interior of appropriate film. A concentration gradient occurs over 2 u.c. for Nd and Sr, revealing an intermixed region of ~2 u.c. (see Fig. S4h in Ref. S1). We note that probe beam broadening and atomic terraces can also lead to a similar profile at an atomically sharp interface. These results thus represent an upper limit for the extent of cation interdiffusion. The same analysis of the rougher STO/NTO interface leads to a mixed region of thickness ~3-4 u.c. Significantly, *these are much smaller than the length scales over which a change in the Ti valence occurs*, establishing that unintentional doping by cation interdiffusion is not responsible for the change in Ti valence, at least outside of the potentially intermixed regions. Additionally, no change in O concentration in excess of the noise occurs across the interface (Fig. S4g in Ref. S1) as would be expected if O vacancies were the driver of interfacial conductivity, or the dominant driver for the observed change in Ti valence across the interface. We also analyzed the oxygen content variation across the STO/NTO/STO interfaces using the STEM-EELS illustrated in Fig. S5 in Ref. S1. The results show that there is no significant change in O concentration across either interface.

**Discussion of the spin-dependent hopping model and its comparison with microscopic phase separation in mixed-valence manganites**

Microscopic phase separation is a common feature of many correlated material systems, including mixed valance manganites, doped cuprate superconductors and others. [e.g. S6 and references therein]. The term describes the coexistence of separate phases in a sample, in a certain parameter range. For example, in manganites, this can involve the appearance of metallic ferromagnetic domains that can vary in size from a few nanometers to hundreds of namometres, with interspersed insulating regions. [S7]. The formation of these spatially-inhomogeneous phases can be due to several factors, including the long-range Coulomb interaction, strain and disorder [S6, S8]. Electronic transport in phase-separated manganite systems exhibiting colossal magneto-resistance is believed to take place via hoping between such metallic ferromagnetic regions, which are embedded in an insulating matrix [S7]. In general, the inhomogeneous phases in these manganites undergo a percolative transition as a function of external parameters such as temperature. Phase separation can also expected in polar-perovskite/$SrTiO_3$ heterostructures due to the large charge densities. This is consistent with our observation of a large negative magneto-resistance and hysteresis in $NdTiO_3$/$SrTiO_3$, which agrees quantitatively with our phenomenological model of spin-dependent hopping between small ferromagnetic regions. We emphasize however that in the absence of a detailed microscopic picture for the formation of these ferromagnetic regions (which is beyond the scope of this work), the similarity between our observations and those in manganites cannot be extended beyond the purely phenomenological level. One marked difference is that we do not observe evidence of a percolative transition, but rather of the gradual blocking of the superparamagnetic behavior of these ferromagnetic regions. This explains why in our study the large negative *MR* and the hysteresis increase upon cooling, in contrast with manganites (where percolation leads to the formation of a connected ferromagnetic phase across the sample, which consequently suppresses spin-misalignment and spin-dependent hopping as samples are cooled below their Curie temperature).